\documentclass[aip, reprint, longbibliography]{revtex4-2}

\draft 
\usepackage{graphicx}
\usepackage{subfigure}
\usepackage{ulem}
\usepackage{color}
\usepackage{bm}
\usepackage{comment}
\usepackage{soul}
\usepackage{amsmath,amssymb}

\usepackage{todonotes}
\definecolor{electricpurple}{rgb}{0.75, 0.0, 1.0}

\newcommand{\MaMi}[1]{\textcolor{black}{#1}}

\begin{document}


\preprint{Submitted to}

\title{Effect of prestress on phononic band gaps induced by inertial amplification}

\author{M. Miniaci}
\affiliation{CNRS, Centrale Lille, ISEN, Univ. Lille, Univ. Valenciennes, UMR 8520 - IEMN, F-59000 Lille, France}
\author{M. Mazzotti}%
\affiliation{Department of Mechanical Engineering, CU Boulder, 1111 Engineering Drive, UCB 427 Boulder, CO 80309, USA}
\author{A. Amendola}
\affiliation{Department of Civil Engineering, University of Salerno, Via Giovanni Paolo II, 132, 84084 Fisciano (SA), Italy}
\author{F. Fraternali}
\affiliation{Department of Civil Engineering, University of Salerno, Via Giovanni Paolo II, 132, 84084 Fisciano (SA), Italy}
\email{marco.miniaci@gmail.com}
\date{\today}
\maketitle 
%
%
\onecolumngrid
\textbf{Phononic crystals and elastic metamaterials have recently received significant attention due to their potential for unconventional wave control.
Despite this interest, one outstanding issue is that their band diagram is typically fixed, once the structure designed.
To overcome this limitation, periodic structures with adaptive elastic properties have recently been proposed for Bragg- and local resonance-driven structures.
\\
In this context, we report about the effect of an applied external mechanical load on a periodic structure exhibiting band gaps induced by inertial amplification mechanism.
If compared to the cases of Bragg scattering and ordinary local resonant metamaterials, we observe here a more remarkable curve shift, modulated through large but fully reversible compression(stretch) of the unit cell, eventually triggering significant (up to two times) enlargement(reduction) of the width of a specific band gap.
An important up(down)-shift of some dispersion branches over specific wavenumber values is also observed, showing that this selective variation may lead to negative group velocities over larger(smaller) wavenumber ranges.
\MaMi{In addition, the possibility for a non-monotone trend of the lower limit of the first BG under the same type of external applied prestrain is found and explained through an analytical model, which unequivocally proves that this behaviour derives from the different unit cell effective mass and stiffness variations as the prestrain level increases.}
These peculiarities derive from the hinge-like behaviour of some regions of the unit cell, which is typical of structures exhibiting the inertial amplification mechanism.
\\
The effect of the prestress on the dispersion diagram is investigated through the development of a 2-step calculation method: first, an Updated Lagrangian scheme, including a static geometrically nonlinear analysis of a representative unit cell undergoing the action of an applied external load is derived, and then the \MaMi{Floquet-Bloch} decomposition is applied to the linearized equations of the acousto-elasticity for the unit cell in the deformed configuration.
\\
Finally, the most evident consequence on the dispersion curves of the application of an external prestress, i.e. the band gap shift with respect to the unloaded structure, is demonstrated through nonlinear transient numerical simulations, clearly proving the capability of the structure to switch from a pass- to a stop-band behaviour over the same frequency range.
\\
The results presented herein provide insights in the behaviour of band gaps induced by inertial amplification, and suggest new opportunities for real-time tunable wave manipulation.}

\section{\noindent \textbf{Introduction}}
Phononic crystals and elastic metamaterials have proved to be powerful platforms to achieve a plethora of unconventional vibrational behaviours deriving from their peculiar dispersion diagrams, characterized by frequency band gaps (BGs) and pass-bands, i.e. frequency regions where the propagation of waves is inhibited or supported, respectively~\cite{deymier2013acoustic}.
This allowed to achieve frequency selective wave filtering~\cite{Hussein_AMR_2014}, guiding~\cite{Semperlotti_PRL, mousavi2015topologically, MiniaciPRX2018} and splitting~\cite{Norris_JAP_2018, Miniaci_PhysRevB_2019}, stimulating the conception of pioneering applications, such as ultra-sensitive devices~\cite{miniaci2017proof}, large scale \MaMi{metamaterials} for vibration shielding~\cite{BrulePRL2014, Miniaci_NJP_2016, fraternali2018innovative}, sub-wavelength imaging~\cite{MoleronNC2015}, elastic wave cloaking / lensing~\cite{ZhangPhysRevLett2011, MisseroniSR2016, PhysRevB_87_174303}, etc.
\\
In this context, an important issue is that phononic crystal and elastic metamaterial operational frequencies are typically fixed once the structure has been designed and fabricated.
In contrast, in the majority of practical applications, including the ones previously mentioned, it may be desirable to dynamically or adaptively tune the BGs (or some of the dispersion branches) in terms of frequency, also after the structure has been fabricated.
To overcome this limitation, periodic systems with adaptive elastic properties have recently been proposed.
For instance, BG tuning has been achieved by means of piezoelectric materials~\cite{Casadei_SMS_2010, Bergamini_ADMA, kherraz2016controlling}, temperature variation~\cite{Jim_APL_2009, Cheng_JASA_2011, Airoldi_NJP_2011, WU2018101}, magneto- and light-based approaches~\cite{Robillard_APL_2009, BouMatar_JAP_2012, GUO201672, ZHANG_PLA_2017, GliozziNatCommPhotoResponsive}, as well as by the application of external mechanical loads~\cite{Bordiga_APL_Prestress}.
Among them, the latter approach can be easily implemented by imposing controlled displacements into specific portions of the structure.
\\
In this context, Bigoni et al.~\cite{BIGONI20082494} proposed for the first time the prestress as a practical way to reversibly alter the dispersion diagram of a periodic structure, including the possibility of shifting the BG position.
The feasibility of the approach was confirmed formulating a theoretical model for an orthotropic, prestressed (compressible) elastic layer vibrating on an elastic half space and assuming long-wave asymptotics for the solution.
Gei et al.~\cite{GEI20103067} relaxed the hypothesis of perfect periodicity and investigated the effect of the prestress in quasi-periodic structures in the case of flexural vibrations.
Amendola et al.~\cite{AMENDOLA201847} studied the band structure of tensegrity mass-spring chains, and the possibility to tune the dispersion relation of such systems by suitably varying local and global prestress variables, given their remarkable softening / stiffening response under axial or compressive loading~\cite{FRATERNALI2015136, AMENDOLA2014234}.
\\
Periodic elastomeric structures, thanks to their capacity of repetitively undergoing large strain deformations in a fully reversible manner, brought to BG nucleation / annihilation mechanisms triggered by the application of external loads radically changing the unit cell geometry~\cite{Babaee_AdvMat_Spirali}.
Deformations in the linear and nonlinear regimes, as well as diverse geometrical topologies have been explored~\cite{Bertoldi_PhysRevB_2008, Bertoldi_PhysRevB_78_184107, Wang_IJSS_2012, Rudykh_PhysRevLett}.
\\
Finally, it has been observed that also mechanical instabilities induced, for instance, by the application of external loads, may alter the propagation of elastic waves.
Slesarenko et al. induced instabilities in soft composite materials achieving a significant decrease of the group velocity (up to going negative) of the transverse waves under specific micropolar conditions~\cite{Slesarenko_APL_2018_NegativeGroupVelocity}.
\\
However, the majority of the aforementioned investigations are limited to the context of (i) Bragg- or (ii) ordinary resonance-induced BGs, whereas in the present paper, we report about the effects of an applied prestress on (iii) periodic structures exhibiting BGs induced by inertial amplification mechanisms.
While in (i) the wave scattering derives from assemblages of periodic unit cells, requiring thus the wavelength of the incoming wave to be comparable to the structural periodicity~\cite{ROMEROGARCIA2013184}, and in (ii) the dynamic behaviour is mainly governed by the eigenfrequencies of resonators included in the structure~\cite{Liu2000}, in (iii) large inertial forces are generated by amplifying the motion of a mass, which in turn increases the inertia of the overall system and lowers its resonance frequency, allowing thus for sub-wavelength and broadband BGs, while keeping the structure lightweight~\cite{Yilmaz2007a, ACAR20136389}.
\\
If compared to the cases of Bragg scattering and ordinary local resonant metamaterials~\cite{MazzottiPrestressFrontiers}, we show that the inertial amplification allows for a more remarkable curve shift, modulated through large but fully reversible compression(stretch) of the unit cell, eventually triggering significant enlargement(reduction) of a specific band gap width.
We also observe a selective up(down)-shift of some dispersion branches, leading to negative group velocities over specific wavenumber ranges.
\MaMi{In addition, a non-monotone trend of the lower limit of the first BG under the same type of external applied prestrain is found and explained through an analytical model, which shows that this behaviour derives from the different unit cell effective mass and stiffness variations as the prestrain level increases.}
Examining the deformed geometries consequent the application of the prestress highlights how the remarkable band shift is due to the hinge-like behaviour of some regions of the unit cell, typical of the inertial amplification mechanism and responsible for a large deformation state, i.e. change of the unit cell geometry.
We consider the static deformation induced by the prestress to be in the linear elastic regime so to have a complete reversibility of the phenomena (tunability).
The analysis is performed in terms of small amplitude motions superimposed on a deformed state once the desired load has been applied.
\\
The paper is organized as follows: in section II, the 2-step Updated Lagrangian scheme, including (i) a static geometrically nonlinear analysis of a representative unit cell undergoing the action of an applied external load and (ii) the \MaMi{Floquet-Bloch} decomposition applied to the linearized equations of the acousto-elasticity for the unit cell in the deformed configuration, is presented.
Section III provides evidence for the dispersion band alteration induced by the application of the prestress in a periodic structure exhibiting BG induced by inertial amplification mechanism.
Parametric dispersion curves for different values of prestress are presented and compared to the original structure (i.e., without the application of any prestress).
Afterwards, a 2-step nonlinear transient numerical simulation confirms the BG shift induced by the prestress in a finite structure, proving its ability to switch from a pass- to a stop-band behaviour over the same frequency range.
The prestress is easily implemented by applying prescribed displacement at specific edges of the structure  prior to the wave propagation.
The full tunability of the structure is guaranteed by the possibility of readily applying and removing the imposed deformations.
Finally, section IV summarizes the main results of the present research and provides future perspectives, shedding light on the possibility of investigating the effect of an additional state of prestress, internal, cross-linking the fields of metamaterials and tensegrity structures.

\section{\noindent \textbf{Floquet-Bloch Analysis of a Prestressed Phononic Crystal}}
In this work, the band structure diagrams are computed using a \MaMi{Floquet-Bloch} finite element method formulated within an Updated Lagrangian scheme.
As schematically presented in Fig.~\ref{UL_Schematics}, the procedure consists of two main computational steps, namely (i) a nonlinear static analysis involving large strains and displacements, and (ii) a so called small-on-large dynamic analysis, in which small vibrations are superimposed on the statically deformed unit cell.
The main computational aspects of the two steps are outlined in the following.

\subsection{Static analysis}
Following the application of a static volume load $\mathbf{f}_{V0}$, or surface load $\mathbf{f}_{S0}$, the unit cell $\Omega_{0}$ identified by the position vector $\mathbf{x}$ and represented by the lattice vector $\mathbf{r}_{0}=\{r_{01},0\}^{\rm{T}}$ and its reciprocal vector $\mathbf{g}_{0}=\{g_{01},0\}^{\rm{T}}=\{r_{01}/2\pi,0\}^{\rm{T}}$, undergoes a displacement $\mathbf{u}_{0}$ that results in a change of configuration from the undeformed state $C_{0}$ to the static deformed state $C$ (see Fig.~\ref{UL_Schematics}).
The deformed domain and matrix of the lattice vectors in the (deformed) configuration $C$ are identified by $\Omega$ and $\mathbf{r}=\{r_{1},0\}^{\rm{T}}$, respectively. The relation between $\mathbf{r}_{0}$ and $\mathbf{r}$ can be expressed as $\mathbf{r} = \mathbf{F}_{L}\mathbf{r}_{0}$, where $\mathbf{F}_{L}$ defines the affine component of the deformation gradient such that $\mathbf{F}_{L}=\mathbf{F}^{-1}\mathbf{F}_{P}$, being $\mathbf{F}=\nabla_{\mathbf{x}}\mathbf{u}_{0}+\mathbf{I}$ the deformation gradient, $\mathbf{F}_{P}$ a periodic non-affine deformation~\cite{zhang2017} and $\nabla_{\mathbf{x}} = \{\partial/\partial x_1, \partial/\partial x_2 \}^{\rm{T}}$.
The equilibrium equations with respect to the undeformed configuration $C_{0}$ can be derived from the variational statement:
\begin{equation}
\int_{\Omega} \left(\mathbf{S}(\mathbf{x}) : \delta\mathbf{E}(\mathbf{x}) - \mathbf{f}_{V0}\cdot\delta\mathbf{u}_{0}\right) da =
\int_{\partial\Omega} \mathbf{f}_{s0} \cdot \delta\mathbf{u}_{0} ds,
\label{eq:var1}    
\end{equation}
\noindent subjected to the Dirichlet boundary conditions:
\begin{equation}
\mathbf{u}_{0}(\mathbf{x}+\mathbf{r}_{0}) = \mathbf{u}_{0}(\mathbf{x}),
\label{eq:DBC1}    
\end{equation}
\noindent in which $\mathbf{E}=\frac{1}{2}(\mathbf{F}^{\rm{T}})\mathbf{F}-\mathbf{I}$ is the Green-Lagrange strain tensor and $\mathbf{S} = \mathbf{D}_{0} : \mathbf{E}$ the second Piola-Kirchhoff stress tensor.
The tensor of tangential elastic moduli, $\mathbf{D}_{0}$, is expressed by $\mathbf{D}_{0} = 4 \partial^2 \psi / (\partial\mathbf{C} \partial\mathbf{C})$, being $\mathbf{C} = \mathbf{F}^{\rm{T}}\mathbf{F}$ the right Cauchy-Green deformation tensor and $\psi$ the elastic energy density. 
The material of the unit cell in $C_{0}$ is specified by the density $\rho_{0}$ while, assuming a hyperelastic material behavior described by the Murnaghan’s model~\cite{murnaghan1937, pau2015, dubuc2017, dubuc2018}, the elastic energy density can be defined as:
\begin{equation}
\begin{split}
\psi =&  \frac{1}{2} \left( \lambda+ 2\mu\right) I_1^2(\mathbf{E}) - 2\mu I_2(\mathbf{E}) + \frac{1}{3}\left( l + 2m \right) I_1^3(\mathbf{E}) \\
& - 2 m I_1(\mathbf{E}) I_2(\mathbf{E}) + n I_3(\mathbf{E}),
\end{split}
\end{equation}
\noindent in which $\lambda$ and $\mu$ denote the first and second Lam\'{e} parameters, respectively, ($l$, $m$, $n$) the third order Murnaghan parameters, and $I_1(\mathbf{E})$, $I_2(\mathbf{E})$ and $I_3(\mathbf{E})$ the first, second and third invariants of the Green-Lagrange strain tensor, respectively. 
\\
The application of a standard Galerkin approach to Eq.~(\ref{eq:var1}) results in the generalized system of equations:
\begin{equation}
\left[ \bm{\Gamma}_{0}^{\mathrm{T}} \mathbf{K}(\mathbf{Q}_{0}) \bm{\Gamma}_{0} \right] \mathbf{Q}_{0}(\mathbf{X}) = \mathbf{P}_{0}(\mathbf{X}),
\label{eq:static2}
\end{equation}
\noindent where $\mathbf{K}(\mathbf{Q}_{0}) $ is the static stiffness matrix, $\mathbf{P}_{0}$ the global vector of nodal forces, $\mathbf{Q}_{0}$ the global vector of independent nodal displacements and $\bm{\Gamma}_{0}$ the mapping operator resulting from Eq.~(\ref{eq:DBC1}) and realizing the condition $\mathbf{U}_{0} = \bm{\Gamma}_{0} \mathbf{Q}_{0}$.
$\mathbf{U}_{0}$ indicates the full vector of nodal displacements.
In this work, the solution of Eq.~(\ref{eq:static2}) is carried out using Comsol Multiphysics 5.3~\cite{comsol}.
\\
Following the Updated Lagrangian scheme, once the displacements $\mathbf{Q}_{0}$ are obtained, the reference configuration is updated from $C_0$ to $C$ by calculating the corresponding nodal coordinates $\mathbf{x}=\mathbf{x}_{0}+\bm{\Gamma}_0 \mathbf{Q}_{0}(\mathbf{x}_{0})$. 
The updated material properties in $C$ are given by $\rho=\rho_0(\det\mathbf{F})^{-1}$ and $D_{ijkl}=(\text{det}\mathbf{F})^{-1}F_{iI}F_{jJ}F_{kK}F_{lL} (D_{0})_{IJKL}$, while the Cauchy stress tensor is obtained from the relation $\bm{\sigma} = (\det \mathbf{F})^{-1} \mathbf{F} \mathbf{S} \mathbf{F}^{\mathrm{T}}$.
The geometry of the unit cell in the configuration $C$ is then re-meshed and used as the basis for the linear dynamic analysis described in the next section.

\subsection{Dynamic Analysis Using the Floquet-Bloch Decomposition}
Following the small-on-large analysis approach~\cite{mazzotti2012, shim2015, pau2015, MazzottiMiniaciBartoliUltrasonics2017, MAZZOTTI2016128}, in which $C$ is the new reference configuration, the position vector for the unit cell in the dynamic deformed configuration $C^{\prime}$ (Fig.~\ref{UL_Schematics}) is approximated as $\mathbf{x}^{\prime}\approx\mathbf{x}$ while, from the application of the Floquet-Bloch theorem, any small harmonic perturbation $\mathbf{u}(\mathbf{x})$ can be expressed as~\cite{collet2011}:
\begin{equation}
\mathbf{u}(\mathbf{x}) = \tilde{\mathbf{u}}(\mathbf{x}) \text{exp}(\text{i}kx) \text{exp}(-\text{i}\omega t),
\label{eq:disp1}
\end{equation}
\noindent in which $\tilde{\mathbf{u}}(\mathbf{x})$ is a $\Omega$-periodic displacement amplitude, $t$ denotes the time, $\omega$ the angular frequency, and $k\in\Lambda$ the Bloch wavenumber, being $\Lambda$ the reciprocal unit cell defined in $C$ by the reciprocal lattice vector $\mathbf{g} = \mathbf{F}_{L}^{-\rm{T}} \mathbf{g}_{0}$.
\\
By defining the $k$-shifted gradient of a generic $\Omega$-periodic vector field $\tilde{\phi}(\mathbf{x})$ as:
\begin{equation}
\nabla_{k} \tilde{\phi}(\mathbf{x}) = \nabla_{\mathbf{x}} \tilde{\phi}(\mathbf{x}) + \mathrm{i} k \tilde{\phi}(\mathbf{x}) \otimes ( \mathbf{r} \| \mathbf{r} \|^{-1} ),
\label{eq:nablak1}
\end{equation}
\noindent the solution of the elastodynamic problem for free vibrations of the unit cell in $C$ subjected to an initial stress $\bm{\sigma}_{0}$ can be obtained from the variational statement:
\begin{equation}
\begin{split}
-\omega^2 & \int_{\Omega} \rho(\mathbf{x}) \delta\tilde{\mathbf{u}}^{*}(\mathbf{x}) \cdot \tilde{\mathbf{u}}(\mathbf{x}) \mathrm{d}\Omega
\\
+
& \int_{\Omega} \delta\tilde{\mathbf{e}}_{k}^{*}(\mathbf{x},\vartheta) : \mathbf{D}(\mathbf{x}) : \tilde{\mathbf{e}}_{k}(\mathbf{x}) \mathrm{d}\Omega
\\
+
& \int_{\Omega} \bm{\sigma}_{0}(\mathbf{x}) : \left[ \left(\nabla_{k}\delta\tilde{\mathbf{u}}^{*}(\mathbf{x})\right)^{\mathrm{T}} \nabla_{k}\tilde{\mathbf{u}}(\mathbf{x}) \right] \mathrm{d}\Omega
= 0,
\end{split}
\label{eq:dyn1}
\end{equation}
\noindent subjected to the Dirichlet boundary condition:
\begin{equation}
\tilde{\mathbf{u}}(\mathbf{x}+\mathbf{r}) = \tilde{\mathbf{u}}(\mathbf{x}) \quad \text{on} \; \partial\Omega,
\label{eq:BCdynamic1}
\end{equation}
\noindent in which $(\cdot)^{*}$ stands for the conjugate of a complex vector or tensor field, $\tilde{\mathbf{e}}_{k}(\mathbf{x}) = \frac{1}{2}[ \nabla_{k}\tilde{\mathbf{u}}(\mathbf{x}) + (\nabla_{k}\tilde{\mathbf{u}}(\mathbf{x}))^{\mathrm{T}}]$ follows from Eq.~(\ref{eq:nablak1}) and denotes the linearized Green-Lagrange strain tensor.
\\
The finite element discretization of Eqs.~(\ref{eq:dyn1}) and (\ref{eq:BCdynamic1}) proceeds by first generating a new mesh for the deformed geometry of the unit cell in $C$ and then applying a Galerkin approach~\cite{mazzotti2014ultrasonic}. As a results, the following generalized linear eigenvalue problem is obtained:
\begin{equation}
\left\{
\bm{\Gamma}^{\mathrm{T}}
\left[
k^2 \mathbf{K}_{3} + \mathrm{i}k \left( \mathbf{K}_{2} - \mathbf{K}_{2}^{\mathrm{T}}\right) + \mathbf{K}_{1} - \omega^{2} \mathbf{M}
\right] 
\bm{\Gamma} \right\}
\tilde{\mathbf{Q}}(\omega)
=
\mathbf{0},
\label{eq:dyn2}
\end{equation} 
\noindent in which $\bm{\Gamma}$ is a mapping operator implementing the Dirichlet boundary condition in Eq.~(\ref{eq:BCdynamic1}), such that $\tilde{\mathbf{U}}=\bm{\Gamma}\tilde{\mathbf{Q}}$, where $\tilde{\mathbf{U}}$ is the global vector of nodal displacement amplitudes and $\tilde{\mathbf{Q}}$ a subvector of $\tilde{\mathbf{U}}$ collecting only its independent components. The expressions of $\mathbf{K}_3$, $\mathbf{K}_2$, $\mathbf{K}_1$ and $\mathbf{M}$ are given in the Appendix A, for the sake of brevity.
\\
The band diagrams of the phononic structure can be computed from the eigenvalue problem in Eq.~(\ref{eq:dyn2}) (i) by extracting the Bloch wavenumbers $k(\omega)$ for any fixed frequency $\omega$, or (ii) by computing the natural frequencies $\omega(k)$ of the system for any fixed Bloch wavenumber $k$.
Since the focus of the present research does not require the knowledge of the spatial attenuation, the latter approach has been used, resulting computationally more convenient, as it does not require the transformation of the system into the state space.

\section{\noindent \textbf{Results}}
To study the effect of an external mechanical load on a periodic structure exhibiting BGs induced by inertial amplification, the numerical method presented in the previous section is here applied to the representative unit cell reported in Fig.~\ref{Model_and_ModeShapes}A.
The unit cell consists of an epoxy matrix exhibiting elongated cross-like holes, and it was proposed for the first time by Acar and Yilmaz~\cite{ACAR20136389}.
Its geometrical and material parameters are reported in Tables~\ref{MaterialProperties},~\ref{GeometricalProperties}.
\begin{table}[]
\caption{Material constants for the epoxy~\cite{MazzottiPrestressFrontiers}.}
\centering
\begin{tabular}{r c c c c c c}
\hline\hline
		& $\rho_0$ [kg$/$m$^3$]	& $\lambda$ [GPa] 	& $\mu$ [GPa] 	& $l$ [GPa] & $m$ [GPa] & $n$ [GPa] \\
\hline
Epoxy 	& 1540 					& 2.59 				& 1.34			& -18.94	& -13.36	& -9.81		 \\
\hline\hline
\end{tabular}
\label{MaterialProperties}
\end{table}
\begin{table}[]
\setlength{\tabcolsep}{3mm}
\caption{Geometrical parameters of the unit cell presented in Fig.~\ref{Model_and_ModeShapes}A.
All the given parameters are in [mm].}
\centering
\begin{tabular}{c c c c c c c c}
\hline\hline
$t_1$	& $t_2$ & $t_3$ & $t_4$	& $d_1$ & $d_2$ & $d_3$ & $d_4$	\\
3.2		& 0.4	& 3.2 	& 0.4	& 5.2	& 2.0	& 20.0 	& 4.0	\\
\hline\hline
\end{tabular}
\label{GeometricalProperties}
\end{table}

\subsection{\noindent \textbf{Band structure analysis}}
As discussed in Section II, the first step for the band structure calculation is to extract the deformation induced in the unit cell by an initial state of stress / strain applied to the structure.
To do this, two different sets of loading conditions are applied to the structure in the form of a normal displacement $ \mathbf{u}_0 (\mathbf{x}_0) \cdot \mathbf{n}_0 (\mathbf{x}_0) $ prescribed to the $x_{02}$-parallel faces of the unit cell (highlighted in purple in Fig.~\ref{Model_and_ModeShapes}A).
The out-of-plane degrees of freedom of the unit cell are blocked, so to prevent any possible bending deformation during the application of the loading conditions.
The first loading set induces a state of compression and varies from $0$ (no prestrained condition) to $ -360 $ $\mu$m (maximum compression condition) with a step of $\Delta \mathbf{u}_0 (\mathbf{x}_0) \cdot \mathbf{n}_0 (\mathbf{x}_0) = - 20$ $\mu$m, while the second set induces a tensile state and varies from $0$ (no prestrained condition) to $ + 130 $ $\mu$m (maximum traction condition) with a step of $\Delta \mathbf{u}_0 (\mathbf{x}_0) \cdot \mathbf{n}_0 (\mathbf{x}_0) = + 10$ $\mu$m.
For the sake of brevity, only the deformation of the unit cell under the conditions of maximum compression / traction are reported in Figs.~\ref{Model_and_ModeShapes}B,C, respectively.
Examining the induced deformation, it is possible to infer that the stress, and thus the deformation, is mainly localized into the hinge-like regions, responsible for the activation of the inertial amplification mechanism~\cite{ACAR20136389}.
The analysis also shows that the maximum Von Mises stress level reached in the structure is of 33.1 and 21.2 MPa, for the two types of loading condition, respectively.
\MaMi{Given these values of maximum stress, and applying the von Mises yield criterion~\cite{mises1913mechanik}, the minimum mono-axial tensile strength required for our material to have a safety factor $\geq 1$ is $\sigma_{min} = 33.1 \cdot \sqrt(3) = 56 $ MPa (if we consider a state of pure shear solicitation), which is an acceptable value for epoxy~\citep{tsiafis2004experimental}.
Considering that the aforementioned value is obtained in the most strict condition of pure shear (which is not the case in the system under consideration), an elastic behaviour of the material, and thus of the band diagram, over the full range $ [ -360, +130 ]$ $\mu$m, modulated by the intensity of the applied prestrain, is guaranteed, allowing for a full reversibility of the undeformed configuration (original dispersion diagram), once the load is removed.
For the sake of completeness, since the small on large theory is considered here, the dynamic component of the stress is negligible with respect to the static prestress.
For this reason it can be assumed that the safety factor does not change between the static and dynamic configuration.}
\\
Once the static analysis performed for the aforementioned sets of prestrain, the deformed geometries are assigned as the input unit cells to calculate the dispersion diagrams exploiting the \MaMi{Floquet-Bloch} theory (see section II).
The out-of-plane displacement of the structure is kept blocked also in this phase, in order to limit the dispersion analysis to waves belonging to the $x_{1} - x_{2}$ plane.
The band structures are computed considering the unit cell to infinitely duplicate in a periodic linear array, and assuming the epoxy in its linear elastic regime (the hypothesis of small displacements is now applied).
The unit cell domain is meshed by means of 8-node hexagonal elements of maximum size $L_{FE} = 0.2 $ mm, which is found to provide accurate eigensolutions up to the frequency of interest~\cite{DEMARCHI2013115}.
It should be noted that, since the applied prestrain applied to each face of the cell induces isotropic deformation in the $x_{1} - x_{2}$ plane in both compression and traction cases, the deformation gradient $\mathbf{F}$ and its affine component $\mathbf{F}_L$ are diagonal and, as a consequence, the orientation of the reciprocal lattice vectors $\mathbf{g}_{1}$ in the deformed configurations does not change with respect to that in the undeformed configuration.
This implies that the orientation of the Bloch wavevector also remains unchanged between the undeformed and deformed configurations (see Fig.~\ref{UL_Schematics}D).
Therefore, the resulting eigenvalue problem $(\mathbf{K}-\omega^2 \mathbf{M})\mathbf{u} = \mathbf{0}$ is solved by varying the non-dimensional wavevector $\textbf{k}^*$ along the irreducible path $ \left[ \Gamma - X \right] $, with $\Gamma \equiv (0, 0) $ and $X \equiv \pi/a, 0)$, being $a = \left( 2 \cdot \sum_{i=1}^3 d_i + d_4 \right)$ the lattice parameter.
\\
Figure~\ref{DispersionCurves} reports parametric plots of the dispersion diagrams of the unit cell as a function of the external prestrain intensity inducing compression (Fig.~\ref{DispersionCurves}A) or traction (Fig.~\ref{DispersionCurves}B) states in the structure.
The dispersion curves are color-coded on the base of the level of prestrain applied at the boundaries of the unit cell in the pre-loading phase.
Specifically, the color bar of Fig.~\ref{DispersionCurves}A varies gradually from $-360$ $\mu$m (blue: maximum compression state) to $0$ (green: unprestrained condition), respectively.
Analysing the band diagrams, it is possible to observe that increasing the compressive state in the unit cell induces (i) a general up-shift (curves shading into dark blue) of the dispersion branches, as well as (ii) group velocity inversion in the third band when $k^*$ gets close to the high symmetry point $X$, as indicated by the black arrow.
The band inversion is more evident comparing the dispersion curves singularly plotted for the $-360$ $\mu$m and $0$ $\mu$m prestrain cases, as reported in Fig.~\ref{FigA_Appendice1}A and Fig.~\ref{FigA_Appendice1}B.
Further inspection of the the dispersion diagram allows to infer that the unprestrained unit cell (green curves) allows for the opening of two BGs in the $[0-3000]$ Hz frequency range, going from $170$ to $470$ Hz and $703$ to $1180$ Hz, respectively (see Fig.~\ref{BGWidthParametrico}A at $0$ $\mu$m imposed displacement).
The upper and lower bounds for both the first and second BGs shift in frequency for increasing values of the prestrain (see Fig.~\ref{BGWidthParametrico}A).
However, although the global width of the first BG experiences a limited variation, both its upper and lower bounds undergo a remarkable frequency shift (they both almost triple their frequency).
On the contrary, the lower bound of the second BG is rather stable in frequency for increasing values of the prestrain, while its upper limit experiences a considerable frequency up-shift.
This is responsible for an overall enlargement of the BG width (almost doubling with respect to the unprestressed case).
Finally, it is possible to notice that the lower bound of the second BG experiences a sort of inflection point when no displacement is imposed.
The different behaviours of the two BGs highlights a selective nature of the prestress in altering the dispersion diagram.
\\
The aforementioned effects partially apply also for the case of unit cell subjected to an external traction pre-loading, i.e. both curve shifting and group velocity inversion are observed (Fig.~\ref{DispersionCurves}B).
However, in this case, group velocity inversion involves more bands (the third and fourth ones) and occurs at lower reduced wavenumber values ($k^* \simeq \frac{\pi}{2a}$), as highlighted by the black arrows in Fig.~\ref{DispersionCurves}B (refer to Fig.~\ref{FigA_Appendice1}B and Fig.~\ref{FigA_Appendice1}C for a direct comparison of the dispersion curves singularly plotted for the $0$ $\mu$m and $+130$ $\mu$m prestrain cases).
This implies that, differently from the compression case, when an external state of traction is induced in the pre-loading phase, only few dispersion curves clearly shift towards higher frequencies (the first, the second, the fifth and sixth bands) over the full range of $k^*$ (from $0$ to $\pi/a$), whereas some others (the third and the fourth ones) exhibit both a down- and an up-shift over a wide range of $k^*$.
As a consequence, while the first BG is almost kept unaltered in terms of frequency width and slightly shifts towards higher frequencies (see Fig.~\ref{BGWidthParametrico}B), the second BG experiences a remarkable width decrease (up to 3 times less the original BG).
\\
\MaMi{At this point, it is worth to point out here the different dynamic behaviour of inertially amplified elastic metamaterials with respect to the ordinary ones when subjected to an external state of prestress that emerges from the present research.
Focusing the attention on the case of prestrain inducing tensile deformation in the structure, in ordinary PCs the BGs tend to decrease its frequency regime, being this effect mainly driven by the geometrical changes of the unit cell rather then by the effective stiffness and mass alteration introduced by the prestress (see for instance Fig. 3A of ref.~\cite{MazzottiPrestressFrontiers} and relevant literature on ordinary PCs presented in section I).
On the contrary, in the case of inertially amplified elastic metamaterials, where the displacement mechanism is used to amplify the effective inertia of small masses~\cite{yilmaz2010theory}, the state of external solicitation represents an important means to control the effective stiffness / effective mass ratio of the structure.
This additional degree of freedom may lead to both an increase or decrease of the lower limit of the first BG frequency regime, under the same type of applied prestrain (see for instance Figs. 3B and 4B, where the lower limit of the first BG increases although a prestress condition inducing tensile solicitation in the structure is applied).
\\
To gain further mechanical insights about this peculiar behaviour, a thorough explanation on the change of the effective stiffness / effective mass ratio of the structure as a function of the applied prestrain is provided with the help of the model reported in Fig.~\ref{FigRev1}, and proposed for the first time by the research group of Yilmaz~\cite{TANIKER2017129, YUKSEL2020138, yuksel2015shape, taniker2015design}.
Figure~\ref{FigRev1}A shows the rigid link equivalent model, which assumes pin joints at the middle points of the flexural hinges.
Lumped parameters $m$, $m_a$ and $k_{eff}$ are reported along with the principal geometrical parameters $\vartheta$ (initial link angle) and displacements ($\delta_x$ and $\delta_y$) consequent the application of a prestrain inducing tensile solicitation in the system.
Already in the undeformed configuration, an angle $\vartheta$ exists between the equivalent rigid links (lines in red) and the ground.
\begin{equation}
f_p = \left( \sqrt{\frac{k_{eff}}{m_{eff}+m}} \right) / (2 \pi) =  \left( \sqrt{\frac{k_{eff}}{m_a(cot^2(\vartheta)+1)/4+m}} \right) / (2 \pi)
\label{Eq_fp_rev3}
\end{equation}
where $m_a(cot^2(\vartheta)+1)/4$ defines the effective mass of the system.
\\
When a tensile prestress is applied, $\vartheta$ decreases to $\vartheta_u$, and the effective mass increases.
If the stiffness $k_{eff}$ is left unaltered, from Eq.~(\ref{Eq_fp_rev3}) it clearly emerges that the lower limit of the first BG would decrease due to increase of the effective mass.
However, if the effective stiffness $k_{eff}$ increases more than the effective mass $m_{eff}$ (the edge mass $m$ does not depend on $\vartheta$), then the lower limit of the first BG may also increase (this reasoning also applies to the case of prestrain inducing a compression state of solicitation and it may lead to analogous considerations).
\\
To verify the correctness of the numerical approach proposed in this paper, a comparison of the analytical values of $f_p$ deriving from Eq.~(\ref{Eq_fp_rev3}) and the numerical solutions is performed and reported in Figs.~\ref{FigRev1}B-D.
Figure~\ref{FigRev1}B reports the parametric plot of the dispersion curves of the unit cell when an external prestrain ($[0-40]$ $\mu$m range) is applied inducing a tensile state in the structure.
The $[100 - 220]$ Hz frequency range is considered.
The same polarization reported in Fig. 3 applies.
The values of the frequency of the longitudinal branch responsible for the lower limit of the first BG are reported at the $X$ high symmetry point in correspondence of the black arrows.
Left panel of Fig.~\ref{FigRev1}C reports the direct comparison of the analytical solution (triangular markers connected by the black line) deriving from Eq.~(\ref{Eq_fp_rev3}) and the numerical results (magenta square markers).
The values of $k_{eff}$ introduced in the analytical model are directly calculated from the deformed numerical configuration according to standard homogenization procedures~\cite{timoshenko2009theory,dirrenberger2019computational}.
Right panel of Fig.~\ref{FigRev1}C shows the variation of the effective stiffness $\Delta k_{eff} $ and of the effective mass $\Delta m_{eff}$.
This explains why the lower limit of the first BG increases, as shown in Figs. 3B and 4B, even if a tensile prestress is applied to the structure.
Indeed, the effective mass increase $\Delta m_{eff}$ (Fig.~\ref{FigRev1}D) is far lower than the stiffness increase $\Delta k_{eff} $ (this is due to the chosen geometrical characteristics of the unit cell, including the out-of-plane dimension $h = t_2$ limiting the value of $m_a$).
As a consequence, it is possible to conclude that in this case the BG alteration is mainly stiffness driven.}
\\
\MaMi{Finally, to clearly show the enhanced potential of inertial amplification PCs with respect to the ordinary ones when an external prestrain condition is applied, additional calculations by changing some of the geometrical and mechanical parameters of the unit cell are performed.
Inspired by the analytical model reported in Fig.~\ref{FigRev1}A, a configuration of the unit cell allowing for the lower limit of the first BG to exhibit a non-monotone trend has been found and reported in Fig.~\ref{FigRev2}.
It is possible to observe that the lower limit of the first BG now first decreases, i.e., the increase of the effective mass is dominant - the red curve is above the blue one at low values of prestrain (yellow rectangle) and then increases, i.e., the increase in effective stiffness is dominant - the red curve is below the blue one at higher values of prestrain (green rectangle).
This has been obtained by changing some of the model parameters as follows: $d_3 = 18.5$ mm, $t_3 = 18.5$ mm, $m_a = 1$ kg$/$m$^3$ and multiplying by a factor of $10$ the Lam\`e and Murnaghan material parameters.}
\\
These results suggest that a deformation of the unit cell geometry induced by a compressive / tensile prestress state, already in the elastic regime, can lead to significant changes in the passband and BG behaviours of a periodic structure, especially if the BG nucleation mechanism is lead by IA.

\subsection{\noindent \textbf{Transient Calculation}}
The confirmation of the above mentioned BG tunability is here verified through a nonlinear transient numerical simulation of wave propagation conducted on a finite waveguide comprising 50 unit cells disposed in the $x_{1}$-direction, as shown in Fig.~\ref{Fig5_Trasmissione_Finale}A.
The idea, here, is to pre-load some unit cells of the waveguide before exciting elastic waves at one of its edge, in order to locally and reversibly change the dispersion diagram and confirm the ability of the waveguide to switch from a pass- to a stop-band behaviour over a specific frequency range.
\\
After having pre-loaded the array, so to uniformly reach the prestrain condition of $-360$ $\mu$m in 6 unit cells (Fig.~\ref{Fig5_Trasmissione_Finale}B), elastic waves are excited at the left edge of the waveguide by means of an imposed displacement of $1$ $\mu$m in the $x_1$-direction (red arrow in Fig.~\ref{Fig5_Trasmissione_Finale}A).
Two input signals are considered: (i) a triangular-like excitation (top-left panel of Fig.~\ref{Fig5_Trasmissione_Finale}C) and (ii) a Hanning modulated 11 sine cycles centered at $1400$ Hz (top-right panel of Fig.~\ref{Fig5_Trasmissione_Finale}C), exhibiting a rather broadband and narrowband frequency content, respectively (bottom panels of Fig.~\ref{Fig5_Trasmissione_Finale}C).
Such pulses have been chosen according to the band structure shifts reported in Fig.~\ref{DispersionCurves} and to highlight the tunable filtering capabilities of the designed waveguide under prestressed conditions.
In both excitation cases, $20$ ms long time transient explicit simulations have been performed in order to allow multiple wave reflections to take place at both the edge of the waveguide and of the prestressed unit cells.
\\
In the case of excitation (i), time transient displacements in the $x_1$-direction are recorded at the two acquisition points R1 and R2 (Fig.~\ref{Fig5_Trasmissione_Finale}B), taken equidistant from the prestressed regions and chosen respectively before (B/f) and after (A/t) the prestressed portion of the waveguide.
After acquisition, signals are Fourier transformed and compared to highlight the differences of the two responses in terms of frequency content.
Figure~\ref{Fig5_Trasmissione_Finale}D reports the displacement along the $x_1$-direction at points R1 and R2 (top panel), as well as their energy content in the frequency domain (lower panel).
The frequencies of the computed BGs as a function of the applied prestrain are also highlighted as a shaded region, where the color scale refers to the level of imposed displacement in the pre-loading phase.
Examining the energy content it emerges that the Fourier transform of the signal acquired before the prestrained region (B/f - red line), where the diagram reported in Fig.~\ref{FigA_Appendice1}B applies, presents component in the $1000 - 1600$ Hz frequency range, whereas the signal registered after the prestrained region (A/t - black line) has no frequency components in this frequency range.
On the contrary, for frequencies above $ \simeq 1600 $ Hz, the amplitudes of the frequency contents are comparable.
This is in agreement with the two different dispersion diagrams and clearly confirms the possibility of the waveguide to to switch from a pass- to a stop-band behaviour over the $1000 -1600$ Hz frequency range by readily applying and removing the imposed deformations, respectively.
\\
When the second type of excitation (i.e., the Hanning modulated 11 sine cycles centered at $1400$ Hz) is applied to the left edge of the waveguide, the switch potential is even more evident.
Indeed, reconstructing the full wave field displacements at specific time instants, it is clearly visible that the pulse is fully supported in the first portion of the waveguide (where no prestress is applied), whereas when the prestrained region begins the wave is strongly reflected back.

\section{\noindent \textbf{Conclusions and future perspectives}}
\begin{figure}
\centering
\begin{minipage}[]{1\linewidth}
{\includegraphics[trim=90mm 45mm 90mm 45mm, clip=true, width=1\textwidth]{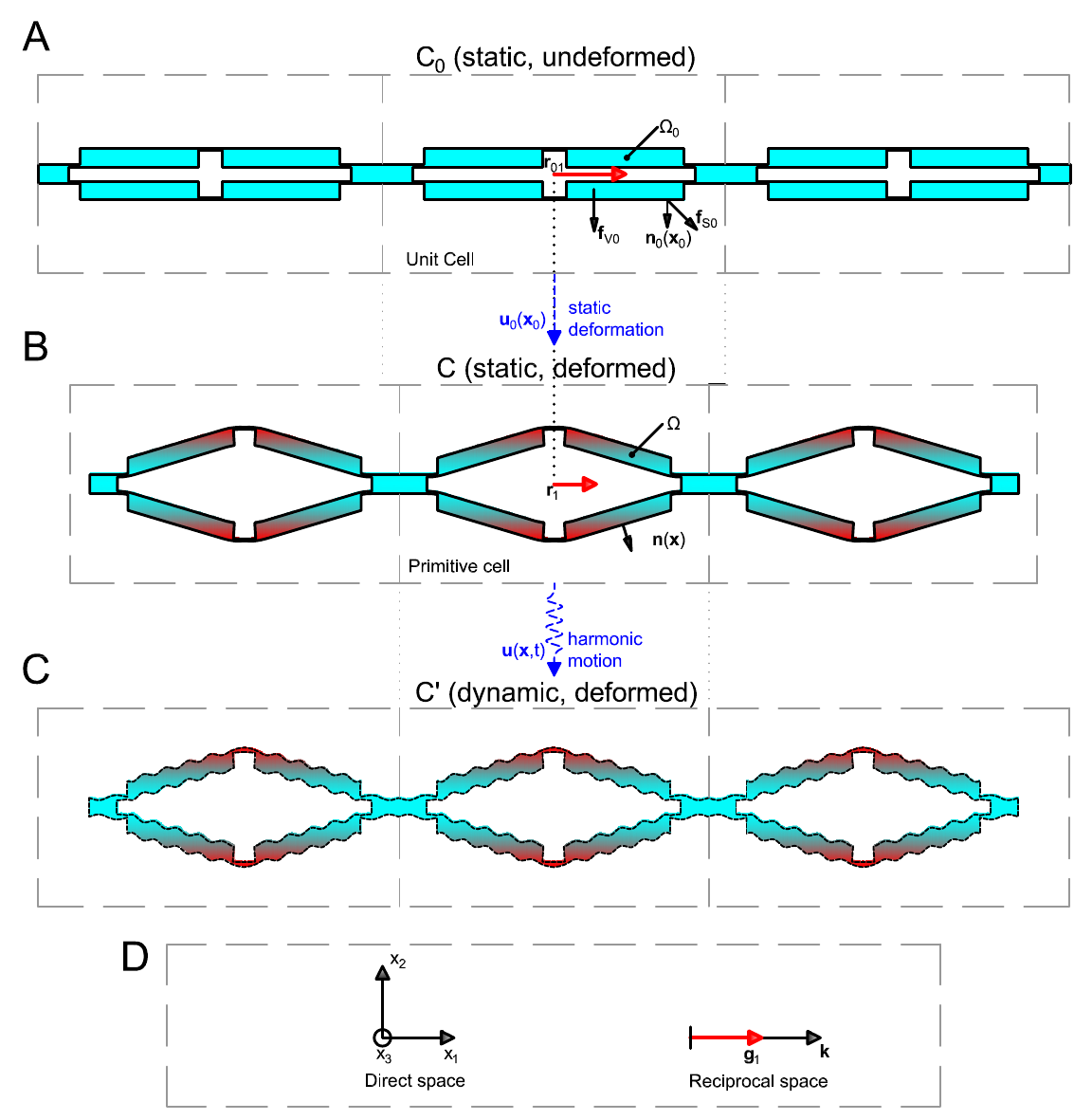}}
\end{minipage}
\caption{\textbf{Graphic representation of the Updated Lagrangian scheme}.
\textbf{(A)} The undeformed configuration $C_0$, i.e. the initial unit cell (delimited by dashed lines) used to calculate the displacement and stress fields introduced by the external mechanical load.
\textbf{(B)} The static deformed configuration $C$, resulting from the application of the external mechanical load.
\textbf{(C)} The dynamic configuration $C'$ undergoing a harmonic motion.
\textbf{(D)} The reference systems in the direct and reciprocal spaces.}
\label{UL_Schematics}
\end{figure}

\begin{figure}
\centering
\begin{minipage}[]{1\linewidth}
{\includegraphics[trim=38mm 59mm 143mm 15mm, clip=true, width=1\textwidth]{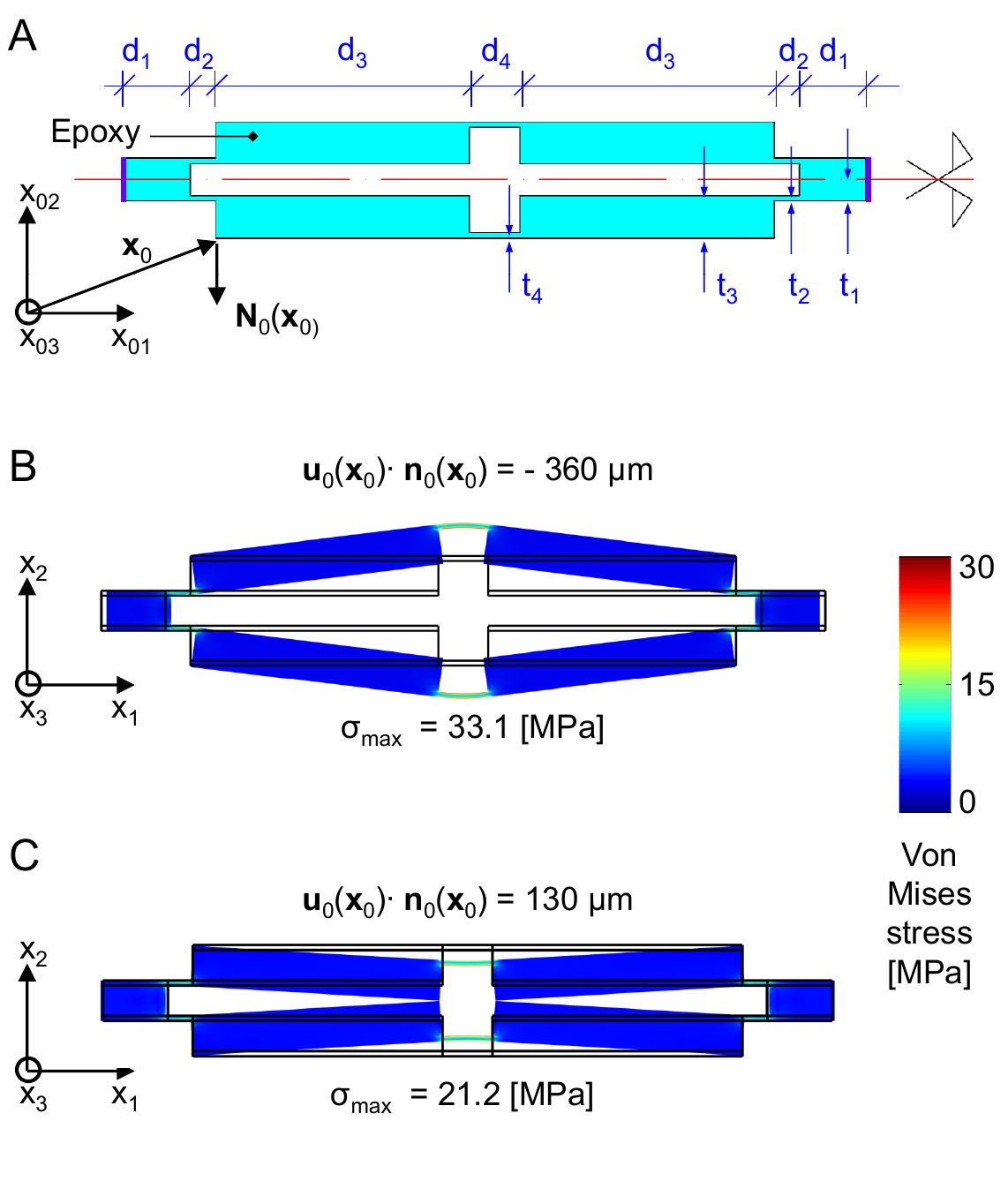}}
\end{minipage}
\caption{\textbf{Schematic representation of the undeformed unit cell and its deformed conditions under isotropic compression / traction.}
\textbf{(A)} Schematic representation of the unit cell exhibiting inertial amplification mechanism, positioned at $\mathbf{x}_0$ with respect to the original reference systems $x_{0i}$, with {i = 1,2,3}.
The structure is made of epoxy and it was proposed for the first time by Yilmaz et al. in ref.~\cite{ACAR20136389}.
\textbf{(B}, \textbf{C)} Deformed configuration of the unit cell under isotropic $\mathbf{u}_0 (\mathbf{x}_0) \cdot \mathbf{n}_0 (\mathbf{n}_0) = 130 \mu$m traction($-360 \mu$m compression).
The colors denote the Von Mises stress in MPa. A maximum stress of 21.2 (\sout{28.6} \MaMi{33.1}) MPa is observed.
Deformations are in 1:1 scale.}
\label{Model_and_ModeShapes}
\end{figure}

\begin{figure}
\centering
\begin{minipage}[]{1\linewidth}
{\includegraphics[trim=64mm 32mm 88mm 79mm, clip=true, width=1\textwidth]{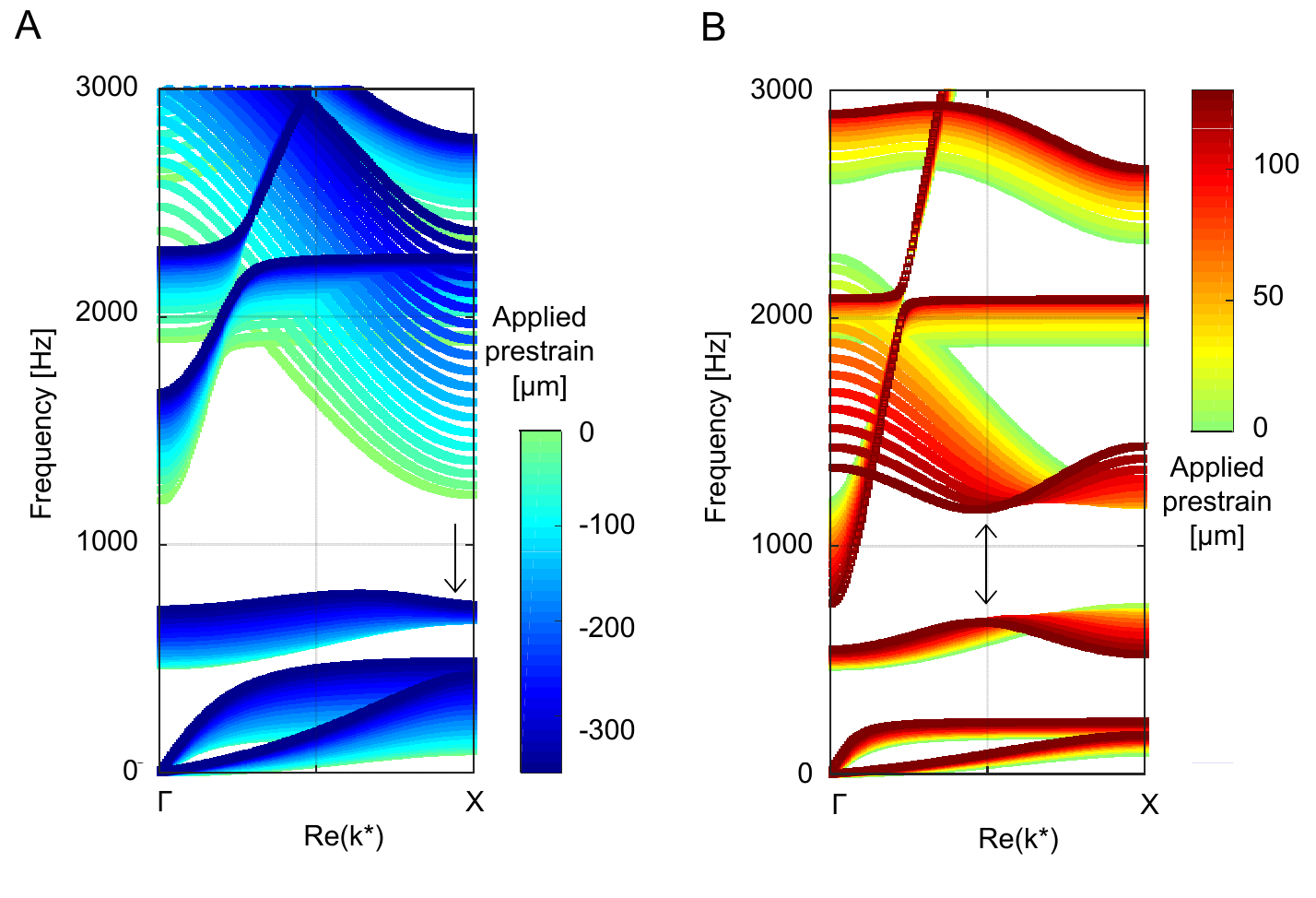}}
\end{minipage}
\caption{\textbf{Band diagrams of the unit cell under different values of applied compression / traction.}
\textbf{(A)} Parametric plot of the the real part of the reduced wavenumber $k^*$ along the $ \Gamma - X $ irreducible path as a function of the applied prestrain inducing a compressive state in the structure.
The dispersion curves are color-coded on the basis of the applied prestrain level at the boundaries of the unit cell in the pre-loading phase.
The polarization factor color bar varies gradually from $-360$ $\mu$m (dark blue: maximum compression) to $0$ (green: unprestrained structure).
A general up-shift trend of the dispersion curves is observable (curves fading into dark blue) and a group velocity inversion (highlighted by the black arrow) occurs for the third band in proximity of the high-symmetry point $X$.
\textbf{(B)} Parametric plot of the the real part of the reduced wavenumber $k^*$ along the $ \Gamma - X $ irreducible path as a function of the applied prestrain inducing a tensile state in the structure.
The dispersion curves are color-coded on the basis of the applied prestrain level at the boundaries of the unit cell in the pre-loading phase.
The polarization factor color bar varies gradually from $130$ $\mu$m (dark red) to $0$ (green: unprestrain structure).
A general down-shift trend of the dispersion curves could be observed (curves fading into dark red).
Group velocity inversions occur, in this case, in more bands (highlighted by the black arrows) and extend over a larger region of $k^*$.}
\label{DispersionCurves}
\end{figure}

\begin{figure}
\centering
\begin{minipage}[]{1\linewidth}
{\includegraphics[trim=64mm 36mm 91mm 79mm, clip=true, width=1\textwidth]{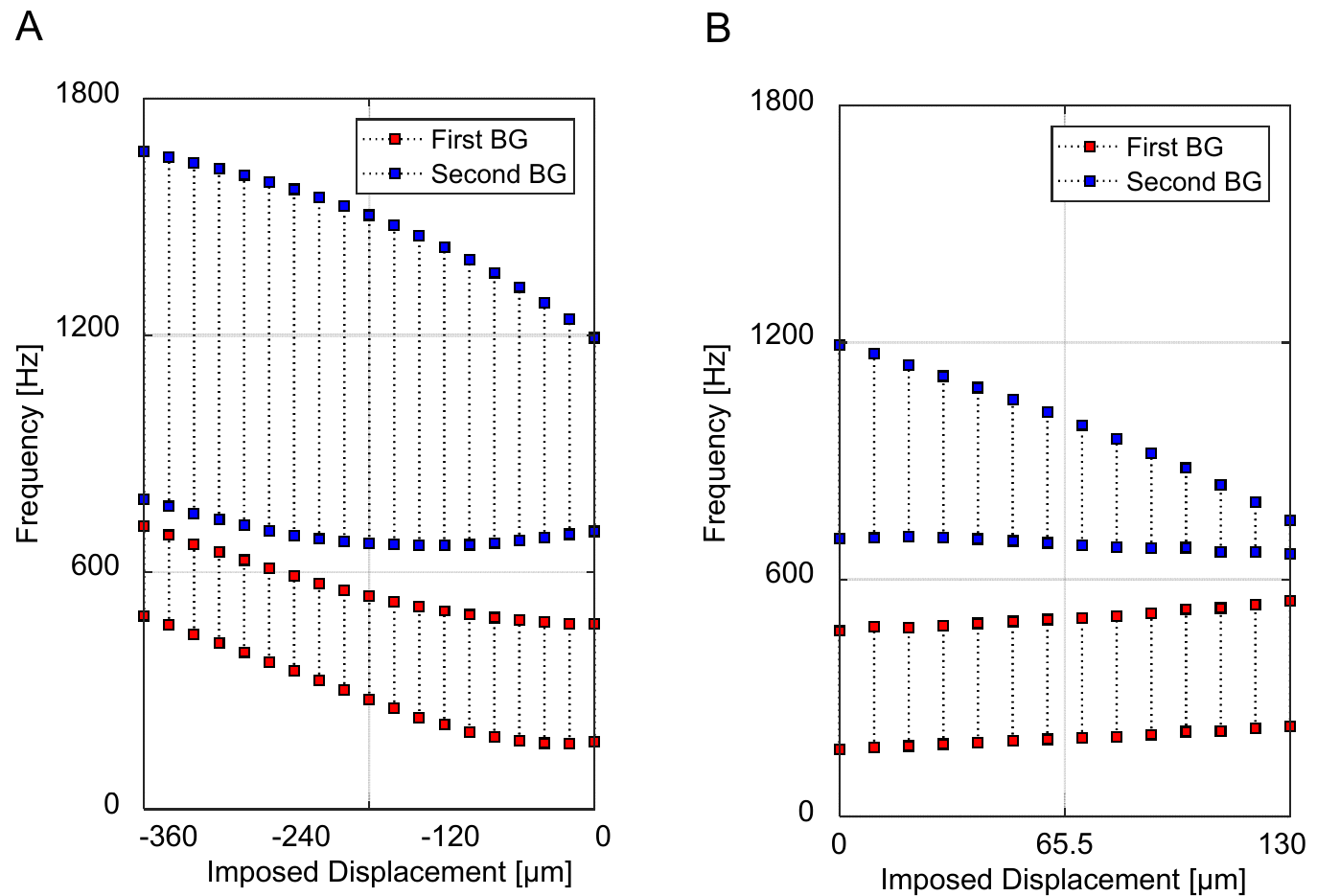}}
\end{minipage}
\caption{\textbf{Upper and lower bounds for the first and second BGs as a function of the applied prestrain inducing a compressive or tensile state in the structure.}
\textbf{(A)} In the case of increasing compressive prestrain, the first BG (black dashed lines delimited by red square markers) experiences a limited width variation, although both its upper and lower bounds are subjected to a remarkable frequency shift (towards higher frequencies).
On the contrary, the lower bound of the second BG (black dashed lines delimited by blue square markers) is rather stable in frequency, while its upper bound still experiences a strong shift towards higher frequencies.
This is responsible for an important global BG width enlargement (twice the initial value).
\textbf{(B)} In the case of prestrain inducing a state of traction in the unit cell, a similar behaviour is observed for the first BG, whereas an important width reduction can be deduced for the second BG (up to three times the initial value), as the initial imposed solicitation is increased to $130$ $\mu$m.}
\label{BGWidthParametrico}
\end{figure}

\begin{figure}
\centering
\begin{minipage}[]{1\linewidth}
{\includegraphics[trim=0mm 45mm 0mm 55mm, clip=true, width=1\textwidth]{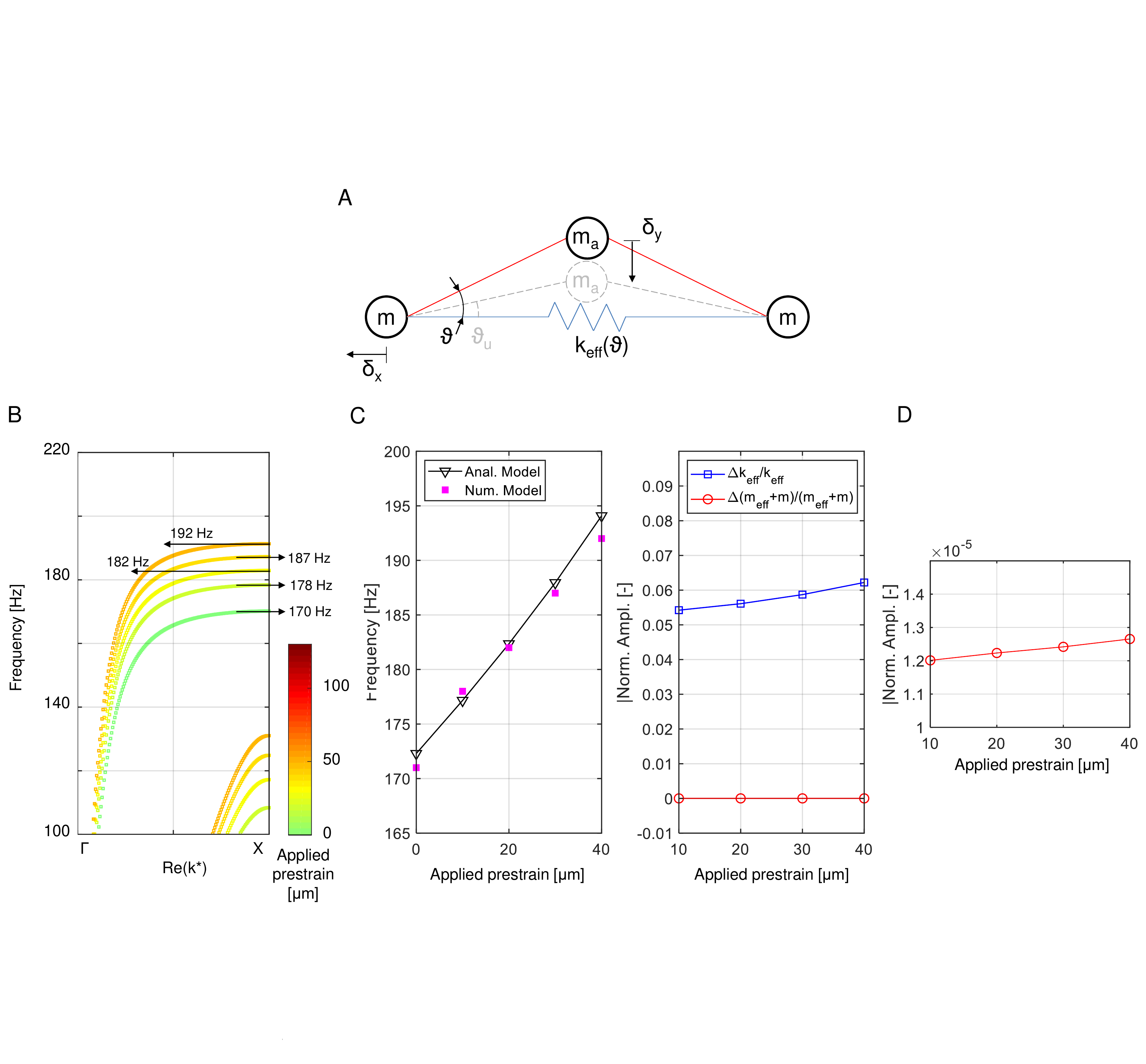}}
\end{minipage}
\caption{\textbf{Change of the effective stiffness / effective mass ratio of the structure as a function of the applied prestrain: numerical - analytical comparison.}
\textbf{(A)} Schematic representation of the rigid link equivalent model of the examined PC with inertial amplification mechanism~\cite{YUKSEL2020138}.
Lumped parameters $m$, $m_a$ and $k_{eff}$ are reported along with the principal geometrical parameters $\vartheta$ (initial link angle) and displacements ($\delta_x$ and $\delta_y$) consequent the application of a prestrain inducing tensile solicitation.
\textbf{(B)} Parametric plot of the the real part of the reduced wavenumber $k^*$ along the $ \Gamma - X $ irreducible path as a function of the applied prestrain inducing a tensile state in the structure in the $[100 - 220]$ Hz frequency range.
The same polarization reported in Fig. 3 applies.
The values of the frequency of the longitudinal branch responsible for the lower limit of the first BG are reported at the $X$ high symmetry point in correspondence of the black arrows.
\textbf{(C)} Comparison of the analytical solution (triangular markers connected by the black line) deriving from Eq. 1 and the numerical values (magenta square markers) is reported in the left panel.
On the right panel, the variation of the effective stiffness $\Delta k_{eff} $ and of the effective mass $\Delta m_{eff}$ explaining the reason of the lower frequency increase of the BG even if a tensile prestress is applied to the structure.
Indeed, the effective mass increase is far lower than the stiffness increase.
\textbf{(D)} Magnification of the change of the effective mass showing its increase.}
\label{FigRev1}
\end{figure}

\begin{figure}
\centering
\begin{minipage}[]{1\linewidth}
{\includegraphics[trim=0mm 70mm 0mm 60mm, clip=true, width=1\textwidth]{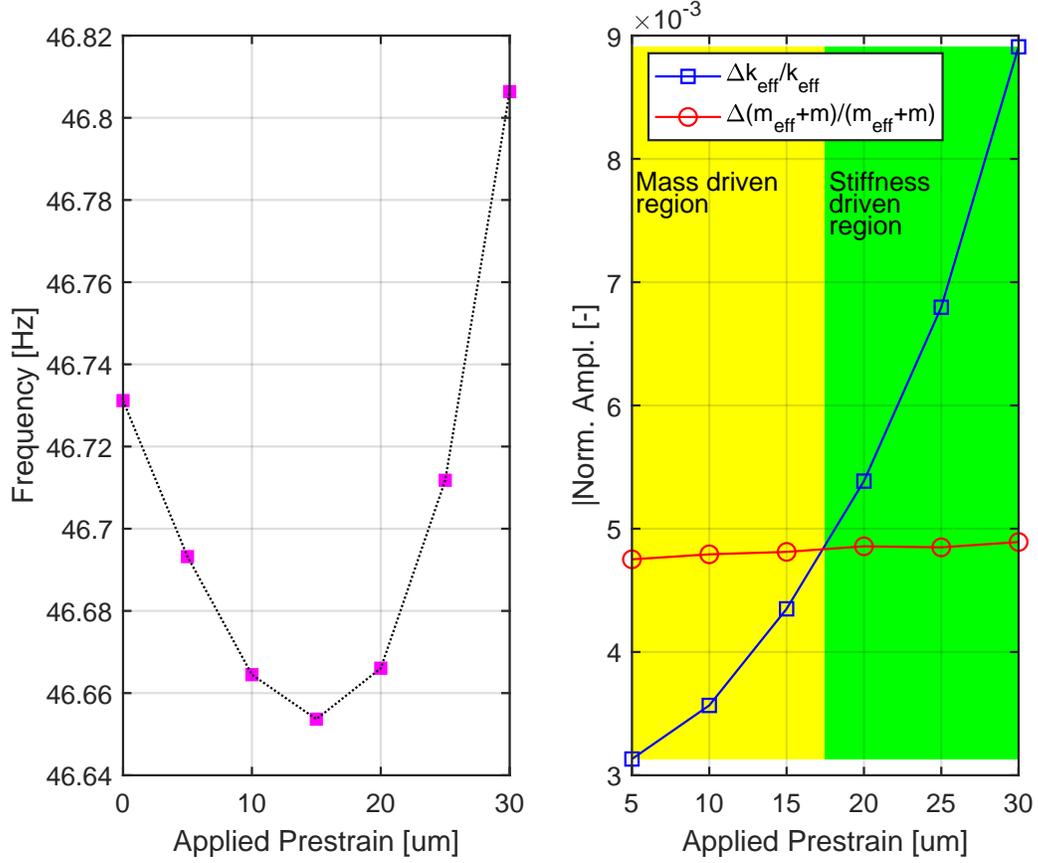}}
\end{minipage}
\caption{\textbf{Additional degree of tunability of inertially amplified elastic metamaterials under external prestrain solicitation state.}
(Left panel) Lower limit frequency of the first BG as a function of the applied prestrain.
(Right panel) The variation of the normalized effective stiffness and of the normalized effective mass explaining the reason of the decrease / increase of the lower limit frequency of the first BG even if the same type of prestress (tension) is applied to the structure.
Yellow and green rectangles determine the mass and stiffness driven regions, respectively.}
\label{FigRev2}
\end{figure}

\begin{figure}
\centering
\begin{minipage}[]{1\linewidth}
{\includegraphics[trim=8mm 58mm 8mm 12mm, clip=true, width=.88\textwidth]{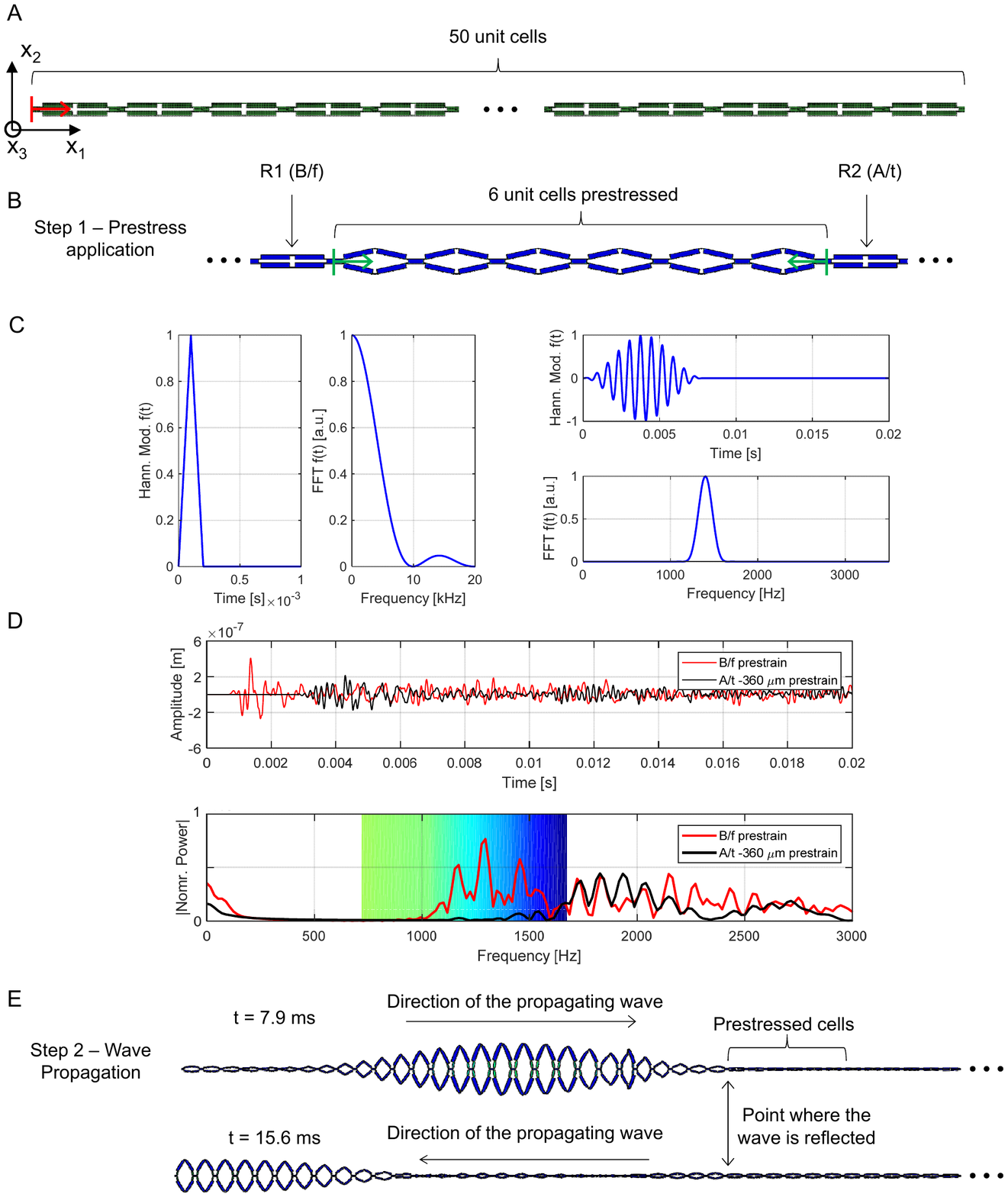}}
\end{minipage}
\caption{\textbf{Switch from a pass- to a stop-band behaviour of the waveguide.}
\textbf{(A)} Schematic representation of the finite structure implemented for the nonlinear transient simulation.
The model comprises 50 unit cells arranged in the $x_{01}$-direction.
\textbf{(B)} Before exciting the propagation of elastic waves, an external load is applied over $6$ internal unit cells of the array, so to locally induce the uniform prestrained condition of $-360$ $\mu$m.
\textbf{(C)} Triangular-like (left panel) and 11 sine cycles centered at $1400$ Hz Hanning modulated (right panel) excitations.
\textbf{(D)} Signals recorded before (B/f) and after (A/t) the $6$ prestressed unit cells (top panel) and their Fourier Transform (low panel), for the triangular-like excitation signal.
The Fourier content clearly shows how the introduction of the prestress alters the frequency response function of the structure.
The tunable BG is reported as a color-coded rectangle.
The green region of the rectangle is the BG corresponding to the unprestrained unit cells (up to $1180$ Hz), whereas the one fading to dark blue corresponds to the BG extension (up to $1680$ Hz) induced by the introduction of the external compressive prestrain (up to $-360$ $\mu$m: dark blue).
The comparison of the two signals proves that the same waveguide is capable to support or to inhibit waves having the same frequency content.
\textbf{(E)} Snapshots at different time steps, for the narrowband excitation signal, showing how the unprestrained waveguide is capable of supporting the propagation of the wave, whereas the stressed region reflect back the majority of the waveform.}
\label{Fig5_Trasmissione_Finale}
\end{figure}

In conclusion, in this work the effect of the application of an external prestress on the dispersion diagram has been investigated.
The geometrical deformation of the unit cell consequent the action of an applied external load has been determined through a static geometrically nonlinear analysis, representing the first step of a so called Updated Lagrangian scheme of calculation.
A \MaMi{Floquet-Bloch} decomposition has then been applied (second step) to the linearized equations of the acousto-elasticity for the unit cell in the deformed configuration.
\\
The effect of the prestress on the the original band structure (i.e., unit cell without any pre-loading phase) has been demonstrated through parametric \MaMi{Floquet-Bloch} analysis and further confirmed by nonlinear transient numerical simulations, proving the capability of the structure to switch from a pass- to a stop-band behaviour in the same frequency range ($[1000 - 1600]$ Hz).
\\
The results presented herein provide insights in the behaviour of band gaps induced by inertial amplification, and suggest new opportunities for real-time tunable wave manipulation.
\\
Future investigations will concern the extension of the present study in the direction of a tensegrity-inspired redesign~\cite{skelton2009tensegrity, fraternali2019mechanical} of the unit cell reported in Fig.~\ref{UL_Schematics}.
A self-similar~\cite{MiniaciPhysRevAppliedGerarchico} \MaMi{reorganization of the structural elements will allow the construction of} a tensegrity architecture.
The design \MaMi{will} derive from the D-bar tensegrity systems~\cite{skelton2009tensegrity, fraternali2019mechanical}, equipped with longitudinal and transverse cables.
The idea \MaMi{will be} to recursively divide the longitudinal span of the cell into segments of equal length \MaMi{and replace} \MaMi{e}ach new segment with a smaller scale unit with equal shape~\cite{skelton2009tensegrity}.
Self-equilibrated tensile forces in the cables and compressive forces in the struts will give rise to an internal state of prestress.
Cables and struts \MaMi{will} present an offset in the transverse direction in order to prevent material overlapping.
\MaMi{T}he application of a self-equilibrated system of forces composed of tensile forces in the cables and compressive forces in the struts will give rise to an \textit{internal} state of prestress, which can superimpose to the external state of prestress, analysed in the present work.
Such initial states of stress will contribute to the geometric term of the stiffness matrix of the structure~\cite{skelton2009tensegrity}, and allow for an extra degree of freedom for the optimal tuning of the dispersion relation of the system~\cite{AMENDOLA201847}.

\vspace*{0.2cm}
\textbf{References}
\bibliography{biblio_IA_prestressed}

\vspace*{0.2cm}
\textbf{Acknowledgements}
MiM has received funding from the European Union's Horizon 2020 FET Open ("BOHEME") under grant agreement No. 863179.

\vspace*{0.2cm}
\textbf{Data availability}.
The data that support the plots within this paper and other findings of this study are available from the corresponding author upon request.

\onecolumngrid 

\newpage
\newpage
\newpage
\newpage
\newpage
\newpage
\newpage

\pagebreak
\clearpage

\section*{\noindent \textbf{Appendix A}}
\label{sec:appA}
The different operators in Eq.~(\ref{eq:dyn2}) are expressed by:
\begin{align}
\mathbf{M} &= \bigcup_{e} \int_{\Omega_{e}} \mathbf{N}^{\mathrm{T}}(\mathbf{x}) \rho(\mathbf{x}) \mathbf{N}(\mathbf{x}) da,
\label{eq:M1}
\\
\mathbf{K}_{1} &= \bigcup_{e}\int_{\Omega_{e}} \mathbf{N}^{\mathrm{T}}(\mathbf{x}) \left[ \mathbf{B}^{\mathrm{T}} \mathbf{D}(\mathbf{x}) \mathbf{B} + \mathbf{B}_{0}^{\mathrm{T}} \bm{\Sigma}_{0}(\mathbf{x}) \mathbf{B}_{0} \right]\mathbf{N}(\mathbf{x}) da
\label{eq:K1}
\\
\mathbf{K}_{2} &= \bigcup_{e} \int_{\Omega_{e}} \mathbf{N}^{\mathrm{T}}(\mathbf{x}) \left[ \mathbf{B}^{\mathrm{T}} \mathbf{D}(\mathbf{x}) \mathbf{H} + \mathbf{B}_{0}^{\mathrm{T}} \bm{\Sigma}_{0}(\mathbf{x}) \mathbf{H}_{0} \right] \mathbf{N}(\mathbf{x}) da,
\label{eq:K2}
\\
\mathbf{K}_{3} &= \bigcup_{e} \int_{\Omega_{e}} \mathbf{N}^{\mathrm{T}}(\mathbf{x}) \left[ \mathbf{H}^{\mathrm{T}} \mathbf{D}(\mathbf{x}) \mathbf{H} + \mathbf{H}_{0}^{\mathrm{T}} \bm{\Sigma}_{0}(\mathbf{x}) \mathbf{H}_{0} \right] \mathbf{N}(\mathbf{x}) \mathrm da,
\label{eq:K3}
\end{align}
\noindent where $\Omega_{e}$ denotes the domain of the $e$-th finite element of the mesh, $\bigcup_{e}(\cdot)$ the standard direct stiffness assembling procedure, $\mathbf{N}(\mathbf{x})$ the matrix of shape functions for the $e$-th element, and $\bm{\Sigma}_{0}(\mathbf{x})$ is a block-diagonal matrix of the form:
\begin{equation}
\bm{\Sigma}_{0}(\mathbf{x}) = \left[
\begin{array}{ccc}
\bm{\sigma}_{0}(\mathbf{x}) & \mathbf{0}                  \\
\mathbf{0}                  & \bm{\sigma}_{0}(\mathbf{x}) \\
\end{array}
\right],
\label{eq:Sigma0}
\end{equation}
\noindent while the different compatibility operators are expressed as:
\begin{align}
\mathbf{B} &= 
\frac{\partial}{\partial x_1}
\left[ 
\begin{array}{cc}
1 & 0 \\
0 & 0 \\
0 & 1 \\
\end{array}
\right]
+
\frac{\partial}{\partial x_2}
\left[ 
\begin{array}{cc}
0 & 0 \\
0 & 1 \\
1 & 0 \\
\end{array}
\right],
\\
\mathbf{B}_{0} &= 
\frac{\partial}{\partial x_1}
\left[ 
\begin{array}{cc}
1 & 0 \\
0 & 0 \\
0 & 1 \\
0 & 0 \\
\end{array}
\right]
+
\frac{\partial}{\partial x_2}
\left[ 
\begin{array}{cc}
0 & 0 \\
1 & 0 \\
0 & 0 \\
0 & 1 \\
\end{array}
\right]
\\
\mathbf{H} &= 
\left[ 
\begin{array}{cc}
1 & 0 \\
0 & 0 \\
0 & 1 \\
\end{array}
\right],
\quad
\mathbf{H}_{0} = 
\left[ 
\begin{array}{cc}
1 & 0 \\
0 & 0 \\
0 & 1 \\
0 & 0 \\
\end{array}
\right].
\end{align}

\section*{\noindent \textbf{Appendix B}}
\label{sec:appB}
In this Appendix a direct comparison of the dispersion curves plotted for the specific cases of $-360$ $\mu$m, $0$ $\mu$m and $+130$ $\mu$m prestrain loads are reported as singular diagrams in Fig.~\ref{FigA_Appendice1}.
\renewcommand{\thefigure}{S\arabic{figure}}
\setcounter{figure}{0}
\begin{figure}
\centering
\begin{minipage}[]{0.95\linewidth}
{\includegraphics[trim=65mm 17mm 58mm 78mm, clip=true, width=1\textwidth]{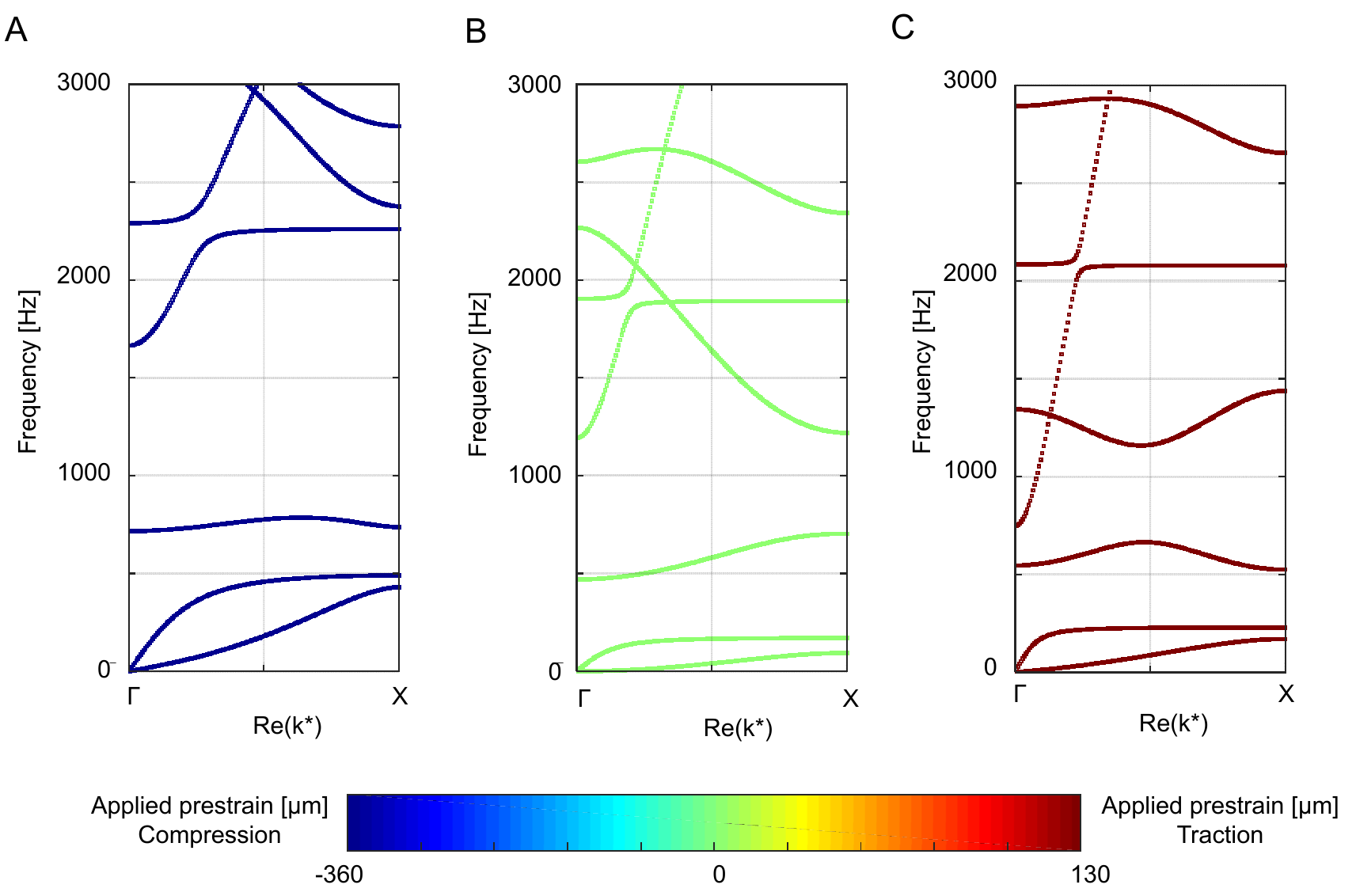}}
\end{minipage}
\caption{\textbf{Plots of dispersion curves for single states of prestress.}
Comparison of the dispersion curves plotted singularly for the (A) $-360$ $\mu$m, (B) $0$ $\mu$m and (C) $+130$ $\mu$m prestrain cases.}
\label{FigA_Appendice1}
\end{figure}

\end{document}